\documentclass[aps,prb,twocolumn,superscriptaddress,amsmath,amssymb]{revtex4-2}

\usepackage{amssymb}
\usepackage{graphicx}
\usepackage{tabularx}
\usepackage{bm}
\usepackage{epsfig}
\usepackage[ansinew]{inputenc}

\begin{document}

\title{Magnetostriction as the origin of the magnetodielectric effect in La$_2$CoMnO$_6$}

\author{M. Boldrin}\thanks{Equal contribution}
\affiliation{Instituto de F\'{\i}sica, Universidade Federal de Goi\'{a}s, 74001-970, Goi\^{a}nia, GO, Brazil}

\author{A. Bagri}\thanks{Equal contribution}
\affiliation{Diamond Light Source, Chilton, Didcot, Oxfordshire, OX11 0DE, United Kingdom}

\author{D. Barlettani}
\affiliation{Diamond Light Source, Chilton, Didcot, Oxfordshire, OX11 0DE, United Kingdom}
\affiliation{London Centre for Nanotechnology and Department of Physics and Astronomy, University College London, London WC1E 6BT, United Kingdom}

\author{E. Teather}
\affiliation{London Centre for Nanotechnology and Department of Physics and Astronomy, University College London, London WC1E 6BT, United Kingdom}
\affiliation{Department of Materials, Imperial College London, London SW7 2AZ, United Kingdom}

\author{L. Squillante}
\affiliation{S\~{a}o Paulo State University (Unesp), IGCE - Physics Department, Rio Claro - SP, Brazil}

\author{M. de Souza}
\affiliation{S\~{a}o Paulo State University (Unesp), IGCE - Physics Department, Rio Claro - SP, Brazil}
\affiliation{S\~{a}o Paulo State University (Unesp), Institute of Theoretical Physics, S\~{a}o Paulo, Brazil}

\author{R. B. Pontes}
\affiliation{Instituto de F\'{\i}sica, Universidade Federal de Goi\'{a}s, 74001-970, Goi\^{a}nia, GO, Brazil}

\author{A. G. Silva}
\affiliation{Instituto de F\'{\i}sica, Universidade Federal de Goi\'{a}s, 74001-970, Goi\^{a}nia, GO, Brazil}

\author{T. J. A. Mori}
\affiliation{Laborat{\'o}rio Nacional de Luz S\'{\i}ncrotron, Centro Nacional de Pesquisa em Energia e Materiais, 13083-970, Campinas, SP, Brazil}

\author{R. Perry}
\affiliation{London Centre for Nanotechnology and Department of Physics and Astronomy, University College London, London WC1E 6BT, United Kingdom}

\author{R. Lora-Serrano}
\affiliation{Universidade Federal de Uberl\^{a}ndia, Instituto de F\'{\i}sica, 38400-902, Uberl\^{a}ndia-MG, Brazil}

\author{E. Granado}
\affiliation{Instituto de F\'{\i}sica ``Gleb Wataghin", UNICAMP, 13083-859, Campinas, SP, Brazil}

\author{E. M. Bittar}
\affiliation{Centro Brasileiro de Pesquisas F\'{\i}sicas, 22290-180, Rio de Janeiro, RJ, Brazil}

\author{L. S. I. Veiga}
\email{larissa.ishibe-veiga@diamond.ac.uk}
\affiliation{Diamond Light Source, Chilton, Didcot, Oxfordshire, OX11 0DE, United Kingdom}

\author{L. Bufai\c{c}al}
\email{lbufaical@ufg.br}
\affiliation{Instituto de F\'{\i}sica, Universidade Federal de Goi\'{a}s, 74001-970, Goi\^{a}nia, GO, Brazil}

\date{\today}

\begin{abstract}

The La$_2$CoMnO$_6$ (LCMO) perovskite has received a lot of attention due to its near room temperature magnetodielectric effect. Despite the recent efforts, the mechanism ruling the correlation between its magnetic and dielectric properties is not yet fully understood. In order to address this issue, we conducted a detailed investigation of the coupling between the structural, electronic and magnetic properties of a polycrystalline LCMO sample. Using magnetic field-dependent x-ray powder diffraction and measurements with a capacitive dilatometer, we show that applying an external magnetic field decreases the unit cell volume, thereby modifying the octahedral distortions. Experiments involving temperature and field-dependent x-ray absorption spectroscopy at the Co-$L_{2,3}$ edges provide further evidence that the spin-orbit interaction of outermost Co 3\textit{d}-orbital and the field-induced enhancement of covalence effects are the key contributors to the magnetostrictive effects. From a detailed analysis using multiplet and density functional theory calculations, we propose that the field-induced modulations of the orbital hybridization and the ligand-to-metal charge transfer are responsible for the changes in the dielectric response of LCMO, thus enabling a direct coupling between magnetic, elastic and dielectric properties in this material.

\end{abstract}

\maketitle

\section{Introduction}

The interest in La$_2$CoMnO$_6$ (LCMO) and La$_2$NiMnO$_6$ (LNMO) perovskites is reminiscent of the early developments of the superexchange theory by P. W. Anderson \cite{Anderson}, J. B. Goodenough \cite{Goodenough1, Goodenough2} and J. Kanamori \cite{Kanamori}, from which both compounds are predicted to be ferromagnetic (FM) insulators due to the positive exchange interaction between half-filled Co$^{2+}$ (3$d^7$ - $t^{5}_{2g}e^{2}_{g}$) or Ni$^{2+}$ (3$d^8$ - $t^{6}_{2g}e^{2}_{g}$) orbitals and Mn$^{4+}$ (3$d^3$ - $t^{3}_{2g}e^{0}_{g}$) orbitals \cite{Blasse}. However, early attempts to produce such compounds, however, have failed to stabilize the transition-metal (TM) ions in the expected oxidation states. Instead, the formation of Mn$^{3+}$ and Ni$^{3+}$ or non-magnetic low-spin (LS) Co$^{3+}$ with the FM character of the samples attributed to vibronic superexchange interactions \cite{Goodenough3} was observed. Over the years, different synthesis routes were used to produce these compounds. Some of them result in nearly Co$^{2+}$/Ni$^{2+}$ + Mn$^{4+}$ FM compounds, while others lead to Co$^{3+}$/Ni$^{3+}$ + Mn$^{3+}$ FM samples, with many materials containing both FM phases \cite{Goodenough4,Goodenough5,PRB2019}.

Recently, the interest in LCMO and LNMO perovskites was renewed after discovering these compounds' near-room-temperature magnetodielectric (MD) effect \cite{Singh,Sarma}. This comes from the attention paid to multiferroism, a term coined to describe materials for which at least two (of the four) ferroic properties - ferromagnetism, ferroelectricity, ferroelasticity and ferrotoroidicity - are present within the same phase. This  opens up possibilities for developing multi-state memory devices, such as electric-field-controlled ferromagnetic resonance devices, transducers with magnetically modulated piezo-electricity, and others.

An important detail about the MD effect on these materials is that an ideal cubic perovskite oxide, $AB$O$_3$ ($A$ = rare-earth/alkaline-earth; $B$ = TM), is centrosymmetric, inhibiting electrical polarization \cite{Hill}. However, structural distortions can lead to an off-center shift of the ions, enabling the realization of polar regions \cite{Hill,Fiebig}. The coexistence of dielectricity and magnetism can also have a nondisplacive character \cite{Fiebig,Liu}. The lone-pair mechanism may drive polar regions where the anisotropic distribution of unbounded valence electrons creates spatial asymmetry \cite{Wang}. It can also be induced by charge ordering, in which the valence electrons are non-uniformly distributed around their host ions in the crystal lattice \cite{Zhu}. This mechanism is particularly relevant in the case of perovskites presenting two distinct TM ions, as is the case of LCMO and LNMO. The polar state can be also directly coupled to the materials' magnetization, as in the exchange-striction mechanism, where symmetric exchange interactions between neighboring spins induce striction along a specific crystallographic direction \cite{Nagaosa}. Finally, another source of MD is the so-called inverse Dzyaloshinskii-Moriya interaction, where an acentric spin structure drives a non-centrosymmetric displacement of charges \cite{Kimura}. 

For LCMO, the polar regions were associated in the first moment with the charge ordering of Co$^{2+}$ and Mn$^{4+}$ \cite{Singh,Lin}. However, larger dielectric constant and MD effect were observed for disordered samples with the presence of Co$^{2+}$/Co$^{3+}$ and Mn$^{4+}$/Mn$^{3+}$ mixed valences \cite{Murthy3}. While some reports  attributed the MD coupling mainly to extrinsic effects such as grain boundaries, antiphase boundaries, defects and contacts \cite{Blasco}, for single-crystalline samples a significant intrinsic contribution is also observed. This is attributed to a small adiabatic polaronic hopping of charge carriers, with the MD effect explained by the spin realignment through an external magnetic field that favors polaronic hopping \cite{Kumar}. An intrinsic contribution is also evident for polycrystalline samples, which appears to be related to the asymmetric hopping mechanism, although a strong influence of the system's microstructure on the dielectric properties is observed \cite{Murthy3}. 

The discussion above evidences that the microscopic mechanism responsible for the MD effect in LCMO is still unclear due to the lack of a systematic and comprehensive study of its origin. In this context, we have employed various techniques to investigate in detail the coupling of the structural, electronic and magnetic properties of polycrystalline LCMO as a function of temperature and external magnetic field. Our results demonstrate a magnetostrictive effect closely related to the MD in this material. We also observe changes in the electron occupation of the 3$d$ orbitals below the magnetic ordering temperature, as a result of stronger ligand-to-metal charge transfer. The presence of a magnetic field ($H$) further increases the metal-ligand orbitals overlap. We discuss our results in terms of the unquenched Co$^{2+}$ orbital moment that strongly interacts with the spin alignment, causing the magnetostriction effects, which in turn affect the material's dielectric response.

\section{Methods}

Polycrystalline LCMO was synthesized by conventional solid state reaction, as described in the Supplementary Material (SM) \cite{SM}. The X-ray diffraction (XRD) data were recorded at different temperatures and $H$ using a Hubber diffractometer in reflection geometry at the XDS beamline at the Brazilian Synchrotron Light Laboratory (LNLS), using a Ge(111) analyzer \cite{XDS}. A double Si(111) crystal monochromator was used to select the incident photon energy $E$ = 19 keV, and a high-throughput LaBr$_3$ scintillator detector in the 2$\theta$ arm was used to collect the data. The Rietveld refinements were performed using the program GSAS+EXPGUI \cite{GSAS}.

The energy-dispersive X-ray spectroscopy was conducted in a JEOL JSM-6610LV low vacuum scanning electron microscope operating with a voltage of 25 kV and a current of 77 $\mu$A. A powdered piece of the LCMO pellet was sprinkled on carbon tape and coated with a 3nm Pt layer to increase signal strength, using a Quorum Q15T ES sputter machine. The experiments were performed in both single point scan and region maps, averaging over many points.

The magnetostriction measurements were performed on a $L_0$ = 0.5 mm long pelletized sample using a quartz capacitive dilatometer cell placed in a Teslatron PT cryostat supplied by Oxford Instruments. A voltage of 0.2 V (15 V) was applied between the dilatometer cell plates for the 150 K and 15 K (300 K) measurements, being $T$ fixed in the mK range and $H$ varied in a 0.1 T/min rate. The sample's length ($L$) variation as a function of $H$, $\Delta L(H)$ = [$L(H)$ $-$ $L(0)$]/$L(0)$, was computed by the relation $\Delta L(H)$ = $\epsilon_0 \pi r^{2}$[$C(H)$ $-$ $C_0$]/$C(H) \cdot C_0$, where $r$ is the radius of the cell's plate, $C$ the measured capacitance and $C_0$ is the initial capacitance value. The dilatometric cell's background was carefully measured and the $H$-dependence of the capacitance without the sample corresponds to 1-2 orders of magnitude lower than the measured values for the investigated sample.

Temperature- and $H$-dependent XAS measurements were conducted at Co, Mn \textit{L}\textsubscript{2,3}- and O \textit{K-}edges at the I06 beamline at Diamond Light Source. We spread fine ground powder of LCMO over conductive carbon tape, and recorded the XAS spectra in total electron yield (TEY) mode. To investigate the effects of magnetostriction in the electronic structure of LCMO, XAS measurements were performed using a 6T/2T/2T superconducting vector magnet at $T =$ 2 K and 300 K and $H$ = 0 and 6 T applied parallel to the incident beam wave vector. To avoid any effect of circular dichroism due to the sample magnetization at low temperatures, we recorded the XAS data with distinct linear polarizations, and took the average data. All the spectra are normalized with ATHENA software and simulated using Crispy software, provided by Quanty \cite{Crispy}. The Co ground state electronic configurations at each temperature and $H$ were estimated in the CTM4DOC software \cite{CTM4DOC}, using the electronic parameters obtained with Crispy.

For the density functional theory (DFT) calculations, we used the projector augmented wave (PAW) method along with the spin-polarized generalized gradient approximation (GGA) as implemented in the Vienna Ab-initio Simulation Package (VASP) \cite{Kresse,Kresse2,Hohenberg,Kohn,Perdew}. In our calculations, a plane-wave cutoff energy of 500 eV was employed. We used the PAW-GGA pseudopotentials (La, Mn, Co, and O) from the VASP distribution with valence configurations of 5s$^2$5p$^6$5d$^1$6s$^2$ for La, 3$d^6$4$s^1$ for Mn, 3$d^8$4$s^1$ for Co, and 2$s^2$2$p^4$ for O \cite{Kresse3}. To sample the k-space Brillouin zone, a 6$\times$6$\times$6 mesh in the Monkhorst-Pack scheme was used \cite{Monkhorst}. All atoms in the system were fully relaxed until the residual forces on the atoms were smaller than 0.01 eV/\AA. To study the properties of LCMO in the presence of chemical disorder, we used a special quasi-random structure (SQS) \cite{Wei}, constructed with the \textit{gensqs} program of the alloy theoretic automated toolkit \cite{Walle}. Such a scheme has been successfully used to simulate random disorder within small periodic supercells convenient for DFT studies \cite{Padilha,Rashid}. The supercells used contained 80 atoms.

\section{Results and discussion}

\subsection{Magnetostriction}
\label{magneto}
  
Polycrystalline LCMO is usually reported as belonging to monoclinic $P2_{1}/n$ or orthorhombic $Pnma$ space group \cite{Goodenough4,Blasco,Fournier}. A fundamental difference between these structures is that for the former, the Co and Mn occupy distinct crystallographic sites, while for the latter, the TM ions share the same site and form a disordered arrangement along the lattice. The difficulty in distinguishing these two crystal structures stems from the very similar scattering factors of Co and Mn for conventional Cu $K_{\alpha}$ wavelength. In this way, the possible ordering of Co/Mn along the lattice is expected to be manifested in very weak Bragg peaks which are usually masked by the background in the conventional XRD experiments. Therefore, additional techniques are necessary to accurately determine the crystal structure \cite{Lin,Fournier,Joy,Joly}. In our case, a previous synchrotron radiation XRD was used to determine that our LCMO sample belongs to $Pnma$ space group \cite{PRB2019}, and this structure is assumed for the subsequent analysis shown in this work. 

\begin{figure}
\centering
\includegraphics[width=1\columnwidth]{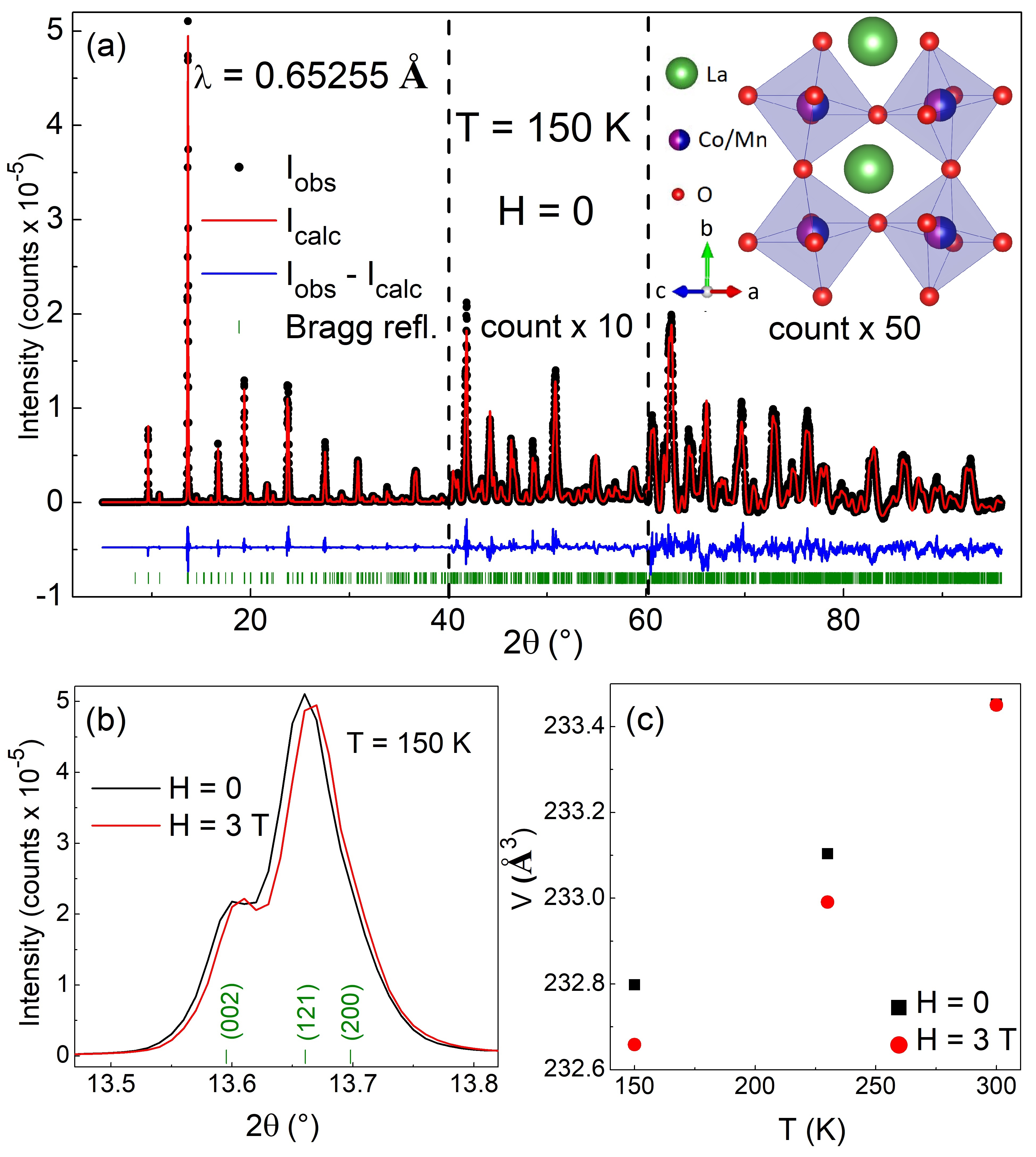}
\caption{(a) Rietveld refinement fitting with the $Pnma$ space group for the XRD pattern measured with incident photons of energy $E$ = 19 keV, at 150 K with $H$ = 0. The intensity is multiplied by 10 in the 40$^{\circ}$ $<$ 2$\theta$ $\leq$ 60$^{\circ}$ range, and by 50 for 2$\theta$ $>$ 60$^{\circ}$, in order to highlight the goodness of fit. Inset shows the crystal structure. (b) Magnified view of the (002), (121), (200) Bragg reflections of the XRD patterns taken at $H$ = 0 and 3 T. (c) Evolution of the unit cell volume with temperature, for $H$ = 0 and 3 T.}
\label{Fig.1}
\end{figure}

Fig. 1 (a) depicts the Rietveld refinement fitting of the XRD pattern obtained at 150 K with $H$ = 0. The main results obtained from the refinement are displayed in Table \ref{T1}. Other diffraction patterns  are shown in the SM, alongside with estimates of the sample's composition conducted by means of EDS \cite{SM}. Fig. \ref{Fig.1}(b) shows a magnified view of the (002), (121), (200) reflections observed at $H$ = 0 and 3 T. As can be seen, the peaks clearly shift towards higher 2$\theta$ values for $H$ = 3 T, indicating a $H$-induced lattice shrinkage. This is evident in Fig. 1(c), which shows the unit cell volume ($V$) obtained from XRD patterns measured at different temperatures and $H$. It is observed that the $H$-induced lattice shrinkage is negligibly small at room temperature, while it gets significantly larger at temperatures close to and below the FM ordering ($T_{C1}\simeq$ 230 K, $T_{C2}\simeq$ 157 K \cite{PRB2019}).

\begin{table*}
    \centering
    \begin{tabular}{c|c|c|c|c|c|c}
    \hline \hline
\multicolumn{7}{c}{Parameters from XRD at 150 K} \\ \hline
    
    $H$  &  $a$ (\AA) & $b$ (\AA) & $c$ (\AA) & $V$ (\AA$^{3}$) & $R_{wp}$ (\%) & $R_p$ (\%) \\ \hline  
    
     0  & 5.4639(1) & 7.7398(1) & 5.5049(1) & 232.80(1) & 11.2 & 8.0 \\ \hline
      
   3 T  & 5.4633(1) & 7.7375(1) & 5.5039(1) & 232.66(1) & 11.4 & 8.9 \\ \hline
   
 & $\Delta a(H)$ = -1.09$\times$10$^{-4}$ & $\Delta b(H)$ = -2.97$\times$10$^{-4}$ & $\Delta c(H)$ = -1.82$\times$10$^{-4}$ & $\Delta V(H)$ = -6.01 $\times$ 10$^{-4}$ \\ \hline
 
 \multicolumn{7}{c}{Bonds and angles} \\ \hline

          &  \multicolumn{2}{c|}{along $a$} &  \multicolumn{2}{|c|}{along $b$} &  \multicolumn{2}{|c}{along $c$} \\ \hline  
    
 &  $H$ = 0 & $H$ = 3 T &  $H$ = 0 & $H$ = 3 T  &  $H$ = 0 & $H$ = 3 T  \\ \hline  

         Co/Mn-O (\AA) &  1.977(6) & 1.968(7) & 1.992(6) & 1.982(7) &  1.957(13) & 1.948(13)   \\ \hline  
         Co-O-Mn ($^{\circ}$) &  156.4(4) &  158.9(4) &  158.3(2) &  161.33(33) &  156.4(4) &  158.9(4) \\ \hline \hline
         
 \multicolumn{7}{c}{Parameters from capacitive dilatometer} \\ \hline 
 
        Temperature:  &  \multicolumn{2}{|c|}{15 K} &  \multicolumn{2}{|c|}{150 K} &  \multicolumn{2}{|c}{300 K} \\ \hline  
        
$H$: & 3 T &  6 T & 3 T & 6 T & 3 T & 6 T \\ \hline    

$\Delta L(H)$: & -3.39$\times$10$^{-4}$  & -4.90$\times$10$^{-4}$  &  -1.87$\times$10$^{-4}$  &  -2.96$\times$10$^{-4}$  & -1.92$\times$10$^{-5}$ & -3.74$\times$10$^{-5}$   \\ \hline \hline          
           
    \end{tabular}
    \caption{Magnetic field effect on the structural parameters of LCMO at 150 K obtained from XRD results, and magnetostriction parameters obtained from a capacitive dilatometer. The numbers in parenthesis in the XRD data represent the standard deviation in the last digit.}
    \label{T1}
\end{table*}

Table \ref{T1} lists the Co/Mn-O bond lengths and Co-O-Mn bond angles obtained from the Rietveld refinements of the XRD patterns carried out at the lowest available temperature, 150 K. It is observed that the bond angle between metal ions and apical oxygen ($b$) is more distorted as compared to the equatorial oxygens ($a$, $c$), indicating anisotropic Co/Mn 3\textit{d}-O 2\textit{p} hybridization. Although care must be taken with these data due to the small O-scattering factor for x rays, they are endorsed by other results that will be discussed later, and agrees with previous studies on LCMO using neutron diffraction and other techniques \cite{Blasco,Bull}. This nearly tetragonal distortion in (Co/Mn)O$_6$ octahedra leads to anisotropic strains along the distinct crystallographic directions with the application of a magnetic field. The $H$-induced strain along the lattice parameter $a$, quantified as $\Delta a(H)$ = ($a_H$ - $a_0$)/$a_0$, where $a_H$ and $a_0$ correspond to the $a$ length at respectively $H$ = 3 T and 0, yields $\Delta a(H)$ = $-$1.09$\times$10$^{-4}$. Analogous calculations along the other lattice parameters show that the shrinkage along $b$ is significantly larger than along $a$ and $c$ (see Table \ref{T1}). Consequently, the more pronounced compression in the Co/Mn bond lengths along the apical direction of the oxygen octahedra results in the straightening of the Co-O-Mn bond angles. 

\begin{figure}
\begin{center}
\includegraphics[width=0.46 \textwidth]{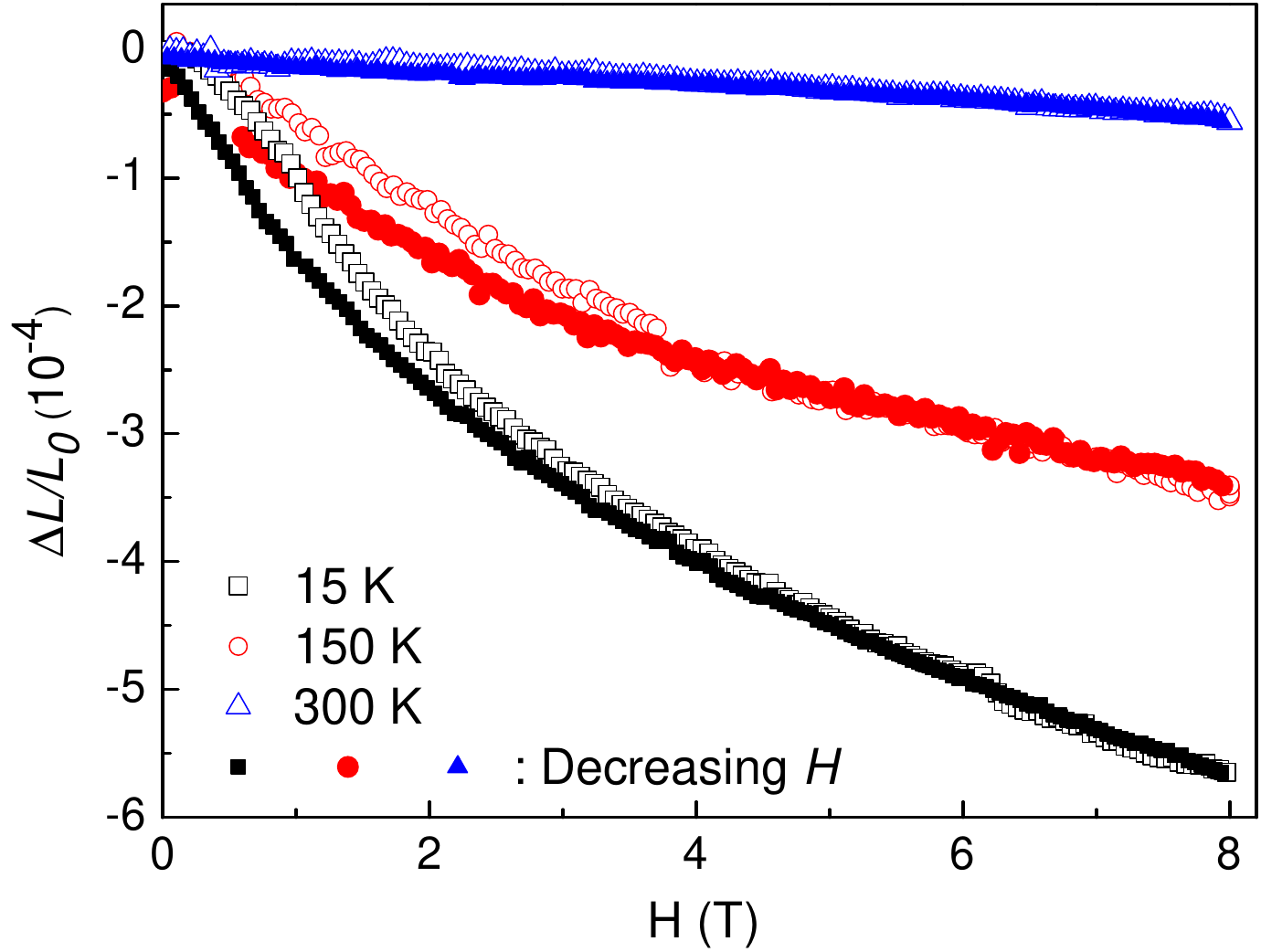}
\end{center}
\caption{Magnetic field dependence of the relative length change at 300 K (blue), 150 K(red) and 15 K (black), measured with a capacitive dilatometer.}
\label{Fig_mgst}
\end{figure}

XRD is not a straightforward method for investigating the magnetostriction of a material. Furthermore,  instrumental limitations precluded the study of $H$-induced structural changes below 150 K. Therefore, we further investigated the magnetoelastic effects of our sample by employing a capacitive dilatometer, with which the temperature could be decreased down to 15 K. The curves obtained at 300, 150 and 15 K, depicted in Fig. \ref{Fig_mgst}, confirm the magnetostriction in LCMO at temperatures below the FM transition, quantified by the relative length change, $\Delta L(H)$. 

Fig. \ref{Fig_mgst} shows hysteretic behavior for the magnetostriction curves measured at 150 K and 15 K, which agrees with previous report \cite{Manikandan}. It suggests the contribution of domain walls on the magnetostrictive effect, but further investigation is necessary to confirm this scenario. One can also notice a decrease in the slope of the 150 K and 15 K curves at high fields, suggesting that a saturation of the magnetostrictive effect  occurs for $H$ $>$ 8 T. More importantly, the $\Delta L(H)$ = -1.87$\times$10$^{-4}$ observed at 150 K for $H$ = 3 T is fairly close to the average of the relative length changes along $a$, $b$ and $c$ obtained from XRD at the same temperature and $H$ (-1.96$\times$10$^{-4}$), supporting the results obtained from powder diffraction. At 15 K, the magnetostriction enhances significantly. However, the $\Delta L(H)$ = $-$4.90$\times$10$^{-4}$ value found for $H$ = 6 T at 15 K is somewhat smaller than that previously reported for polycrystalline LCMO ($\sim$ $-$6.30$\times$10$^{-4}$ at 10 K, $H$ = 5 T \cite{Manikandan}). Besides the distinct temperatures, such discrepancy may be attributed to the difference in the Co$^{2+}$/Co$^{3+}$ ratio of these samples. As discussed later, the magnetostriction effect in LCMO is closely related to the spin-orbit interaction in Co$^{2+}$ and, albeit both samples were produced by solid state reaction, distinct synthesis temperatures were employed \cite{PRB2019,Manikandan}. It was already shown that the Co$^{2+}$/Co$^{3+}$ proportion in LCMO is particularly sensitive to the synthesis condition \cite{Goodenough4}.

\subsection{X-ray absorption spectroscopy}

\paragraph{\textbf{Charge states:}}

\begin{figure}
        \centering
        \includegraphics[width=1\columnwidth]{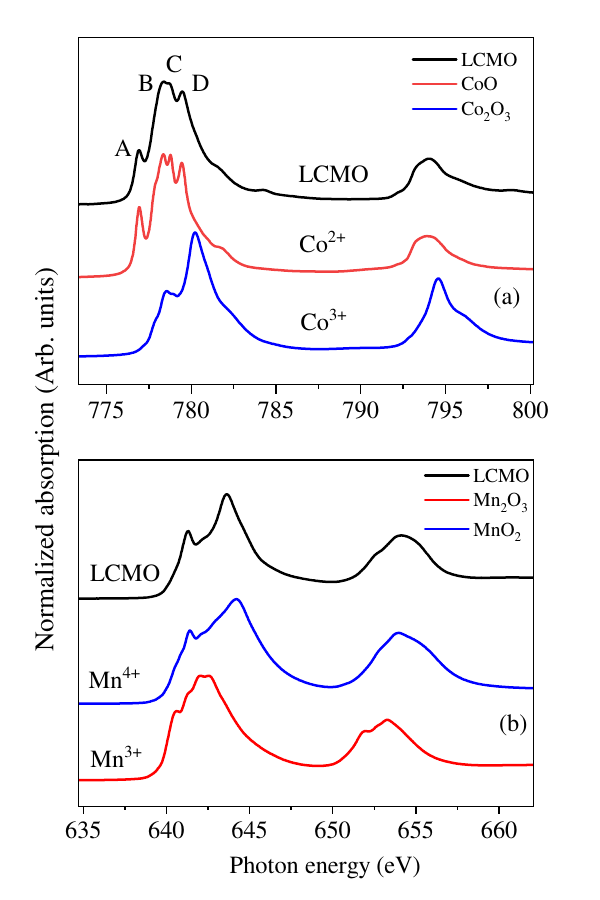}
        \caption{Room temperature XAS spectra of LCMO at (a) Co and (b) Mn $L$-edges with the reference samples.}
        \label{fig:2}
    \end{figure}

To determine the Co and Mn valence states in LCMO, we have compared the recorded XAS spectra at Co and Mn $L_{2,3}$-edge with that of reference samples. Fig. \ref{fig:2}(a) shows the  Co $L_{2,3}$-edge XAS spectra for LCMO alongside Co$^{2+}$ (CoO) and Co$^{3+}$ (Co$_2$O$_3$) references. The observed Co $L_{2,3}$-edge XAS of LCMO is dominated by multiplet structures arising due to the electronic transitions from the $^4F_{9/2}$ ground state to $^4G$, $^4F$, $^4D$, $^2G$, $^2F$ final states, as shown at Co \textit{L}\textsubscript{3}-edge, comprising of features A ($\sim$776.9 eV), B ($\sim$778.3 eV), C ($\sim$778.7 eV) and D ($\sim$779.6 eV). The similar multiplet structures and matching photon energy position of Co $L_3$-edge of LCMO and CoO, confirms a majority Co$^{2+}$ valence state. However, the small shoulder at the higher photon energy side of the $L_3$-edge indicates a marginal presence of Co$^{3+}$ \cite{mir}. Previous studies revealed that Co$^{2+}$ stabilizes in high spin (HS) ground state in LCMO, with mainly $t_{2g}$$^{5}e_g$$^2$ ($S$ = 3/2) electronic configuration, while Co$^{3+}$ stabilizes in non-magnetic low spin (LS) state with $t_{2g}$$^{6}e_g$$^0$ ($S$ = 0) electronic configuration at low temperature \cite{PRB2019}.

In Fig. \ref{fig:2}(b), Mn $L_{2,3}$-edge of LCMO is compared with the Mn$_2$O$_3$ (Mn$^{3+}$) and MnO$_2$ (Mn$^{4+}$) reference samples. The line-shape of Mn the \textit{L}-edge of LCMO is similar to that of MnO$_2$, confirming the expected dominant presence of Mn$^{4+}$ ionic state. However, the energy position of the \textit{center of gravity} of  Mn $L_3$ peak is found to be in-between those of Mn$^{3+}$ and Mn$^{4+}$, suggesting a minor presence of Mn$^{3+}$ charge state. The mixed valence character of LCMO is usually associated to the presence of anti-site disorder, resulting in competing magnetic phases where, besides the Co$^{2+}$-O-Mn$^{4+}$ coupling, other exchange interactions such as Co$^{3+}$-O-Mn$^{3+}$, Co$^{2+}$-O-Co$^{3+}$ , Mn$^{3+}$-O-Mn$^{4+}$, Co$^{2+}$-O-Co$^{2+}$, Mn$^{4+}$-O-Mn$^{4+}$ may play a role. In the case of our LCMO sample, the presence of both Co$^{2+}$-O-Mn$^{4+}$ and Co$^{3+}$-O-Mn$^{3+}$ FM interactions is revealed by the two transition temperatures, $T_{C1}$ = 230 K and $T_{C2}$ = 157 K \cite{PRB2019}.

\begin{figure}
\begin{center}
\includegraphics[width=0.48 \textwidth]{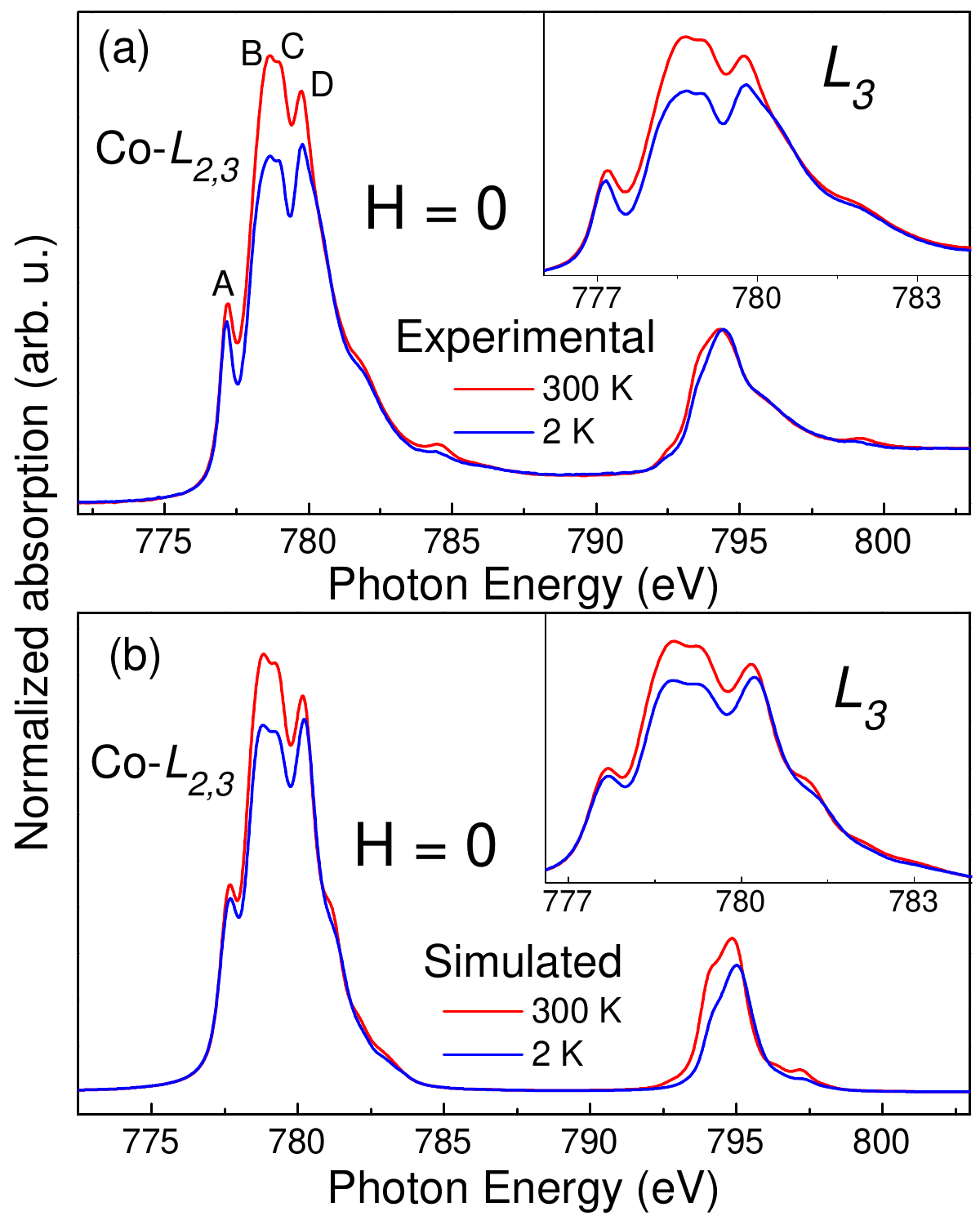}
\end{center}
\caption{(a) Experimental and (b) simulated Co $L_{2,3}$-edge spectra of LCMO at 300 K and 2 K. The insets show magnified views of the temperature-induced modulation of features B, C and D.}
\label{Fig_CoL_T}
\end{figure}

\paragraph{\textbf{Temperature-dependent electronic structure:}}

Before investigating the role of $H$ on the electronic structure of LCMO, we turn our attention to how it gets modified by temperature, in the absence of an external magnetic field. The expected compression of the unit cell volume with decreased temperature alters the cation-ligand bond lengths and angles, leading to changes in the Co/Mn 3\textit{d}-O 2\textit{p} orbital overlap. This gives rise to changes in the electronic parameters such as the on-site Coulomb repulsion ($U_{dd}$), charge transfer energy ($\Delta$\textsubscript{CT}), crystal field (10Dq) and metal-ligand hybridization, which affects the XAS spectra. 

In the main panel of Fig. \ref{Fig_CoL_T}(a), we compare the normalized Co \textit{L}-edge XAS spectra at 300 K and 2 K. It is observed an overall decrease of the spectral weight at low temperature, suggesting an increased electron population of Co site resulting from the compression-induced enhancement of the Co 3$d$ - O 2$p$ orbital hybridization. Furthermore, it may be noticed a relative increase of feature D (see the inset). Here we recall that the electronic structure of octahedrally coordinated Co results from a delicate balance between the crystal field and interatomic exchange interactions, leading to the low-spin (LS), high spin (HS) or even intermediate spin (IS) configurations in the same perovskite \cite{Asai, Raveau}. The changes observed in the relative intensities of the multiplet features are likely associated with the strengthened crystal field, orbital hybridization, and charge transfer due to the closer Co-O bonds, resulting in an internal redistribution of electron occupancy in the Co-3$d$ orbitals.

\begin{table}
\caption{Electronic parameters for the Co $L$-edge XAS simulations performed at distinct temperatures and $H$, and the corresponding ground-state configurations for Co.}
\label{tab:2}
\begin{tabular}{ccccccc}
\hline \hline
Condition: & & $T$ = 300 K & & $T$ = 2 K & & $T$ = 2 K \\
 & & $H$ = 0 & & $H$ = 0 & & $H$ = 6 T \\ 
\hline

$F_k$ & & 0.77 & &  0.72 & &  0.73 \\

$G_k$ & & 0.75 & &  0.73 & &  0.73 \\

$L$ FWHM & & 0.42 & &  0.42 & &  0.40 \\

$G$ FWHM & & 0.42 & &  0.42 & &  0.40 \\
 
10Dq & & 0.95 & &  1.01 & &  1.05 \\

$\zeta_{3d}$ & & 0.07 & &  0.1 & &  0.13 \\

$\Delta_{CT}$ & & 1.70 & &  0.60 & &  0.49 \\

$U_{dd}$ & & 4.2 & &  4.2 & &  4.2 \\

$U_{pd}$ & & 5.6 & &  4.3 & &  4.2 \\

$V(e_g)$ & & 0.35 & &  0.45 & &  0.55 \\

$V(t_{2g})$ & & 0.30 & &  0.35 & &  0.45 \\

$n$(3$d$) & & 7.03  & &  7.11 & &  7.15 \\

$H_{ex}$($x,y,z$) & & 0  & &  0.001 & &  0.01 \\

$B_{ex}$($x,y,z$) & & 0  & &  0 & &  0.001 \\

\hline
\multicolumn{7}{c}{Co ground-state electronic configurations*:} \\

$t_{2g}$$^{5}e_g$$^{2}$ (\%) & & 87.42 & & 83.91  & & 81.04 \\

$t_{2g}$$^{4}e_g$$^{3}$ (\%) & & 10.23 & & 9.01  & & 8.27 \\

$t_{2g}$$^{6}e_g$$^{2}\underline{L}$ (\%) & & 1.31 & &  5.12 & & 8.03 \\

$t_{2g}$$^{5}e_g$$^{3}\underline{L}$  (\%) & & 0.97 & &  1.84 & & 2.49 \\

$t_{2g}$$^{4}e_g$$^{4}\underline{L}$ (\%) & & 0.07 & &  0.12 & & 0.16 \\

\hline \hline
\multicolumn{7}{c}{\footnotesize *$\underline{L}$ denotes a hole at O 2$p$.}
\end{tabular}
\end{table}

To further understand the influence of temperature on the Co \textit{L}-edge spectra, we have simulated the Co\textsuperscript{2+} \textit{L}-edge at 300 K and 2 K using the semi-empirical multiplet approaches in Crispy software \cite{Crispy}. The lineshape of Co \textit{L-}edge is very sensitive to 10Dq and $\Delta$\textsubscript{CT}, along with the ground and excited states, which can be controlled by the two-particle interaction parameters. Therefore, we have performed charge-transfer multiplet calculations by varying the Slater integrals, the 3$d$ spin-orbit coupling ($\zeta_{3d}$), $U_{dd}$ and $U_{pd}$, $\Delta$\textsubscript{CT}, 10Dq, and the O 2\textit{p} - Co 3\textit{d} hybridization strength, while the other parameters were kept fixed at the Crispy's standard values for Co$^{2+}$. The simulated Co$^{2+}$ $L_3$-edge spectra are shown in Fig. \ref{Fig_CoL_T}(b), evidencing a good match with the experimental spectra.

The values of the parameters used for the simulations are listed in Table \ref{tab:2}. It is observed the expected increase of 10Dq with lowering temperature, related to the lattice shrinkage. On the other hand, $\Delta$\textsubscript{CT} substantially decreases by 1.1 eV at 2 K compared to RT, suggesting the increased number of 3\textit{d}-electrons at Co site due to enhanced covalency, as revealed by the decreased intensity of the Co XAS spectrum. The fact that $U_{dd}$ $>$ $\Delta$\textsubscript{CT} at both temperatures suggests the positive charge transfer insulator character of LCMO in the Zaanen$-$ Sawatzky$-$ Allen (ZSA) diagram \cite{zaanen}. 

The scenario drawn above is supported by other parameters, such as the hopping parameters $V$ that define the hybridization between Co 3$d$ and O 2$p$ orbitals. $V(e_g)$ and $V(t_{2g})$ are related to the Slater-Koster parameters, \textit{pd$\sigma$} and \textit{pd$\pi$}, by the relation \textit{pd$\pi$}$\sim$0.5\textit{pd$\sigma$}. It is important to mention that the overlap integral determines the hybridization strength \textit{pd}$\sigma$ between ligand and metal orbitals and mostly depends on the metal-ligand distance, bond angle and ionic radii, given as \textit{pd}$\sigma$ = cos($\pi$-$\beta$)[r\textsuperscript{1.5}/d\textsuperscript{3.5}] \cite{Harrison,Jana,Bagri}. Both $V(e_g)$ and $V(t_{2g})$ enhance with decreased temperature. Therefore, the reduction of the bond lengths caused by the lattice shrinkage with lowering temperature leads to an enhancement of Co 3\textit{d} - O 2\textit{p} orbital overlap, thereupon increasing the electron population at Co orbitals that is manifested in the decreased spectral weight in the XAS data. Contrasting to the Co $L$-edge XAS data, the changes observed at the Mn $L_{2,3}$-edge are negligibly small. Further details can be found in SM \cite{SM}.

\paragraph{\textbf{Magnetic field-dependent XAS study:}}
  
\begin{figure}
\begin{center}
\includegraphics[width=0.48 \textwidth]{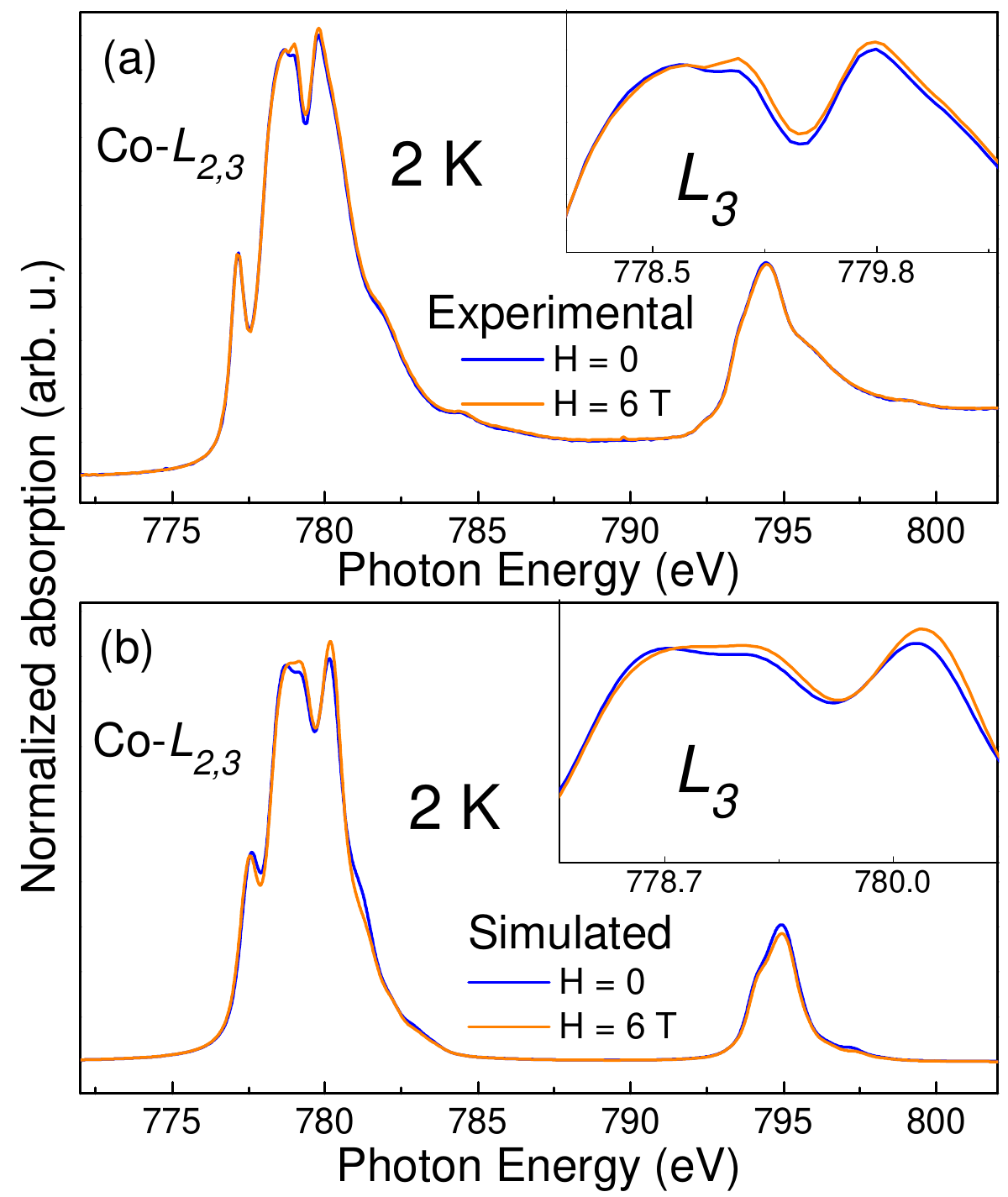}
\end{center}
\caption{(a) Experimental and (b) simulated Co $L_{2,3}$-edge spectra of LCMO at 2 K, measured with $H$ = 0 and 6 T. The insets show magnified views of the $H$-induced modulation of $L_{3}$-edge.}
\label{Fig_CoL_H}
\end{figure}

Our magnetostriction studies suggest that the $H$-induced lattice shrinkage affecting the TM 3$d$ - O 2$p$ hybridization is expected to play a role on the Co-O-Mn exchange interactions. To verify such changes, we have recorded $H$-dependent XAS spectra on LCMO at Co and Mn $L_{2,3}$- and O \textit{K}-edge. Fig. \ref{Fig_CoL_H}(a) shows the normalized XAS spectra of Co \textit{L}-edge carried out at 2 K with $H$ = 0 and 6 T with average polarization. 

The multiplet structure of Co $L_{3}$-edge shows significant modulation with $H$, as highlighted in the inset of Fig. \ref{Fig_CoL_H}(a). Similarly to the effect of decreasing temperature, $H$ acts to decrease the relative intensities of features A, B and D. However, a closer inspection reveals that the spectral weight of feature C increases, contrasting with the effect observed when temperature is reduced. This indicates additional $H$-induced modulation of the Co 3$d$ electronic structure. Our DFT studies discussed in Section \ref{DFT} suggest the increase in the density of unoccupied states on $d_{yz}$ orbital with the application of $H$, which may be the main reason behind the observed $H$-induced increase in the intensity of feature C. We shall return to this subject in the following sections.

To further examine the interplay between $U_{dd}$, $\Delta$\textsubscript{CT}, 10 Dq and superexchange interaction in presence of $H$, we have simulated the Co$^{2+}$ $L$-edge XAS at 2 K for $H$ = 6 T, taking the contribution from magnetic exchange interactions. A comparison between the $H$ = 0 and $H$ = 6 T simulated spectra is depicted in Fig. \ref{Fig_CoL_H}(b). The estimated electronic parameters used for the simulations are shown in table \ref{tab:2}, where a slight increase of 10 Dq is observed when $H$ is applied, which agrees with the magnetostrictive character of LCMO. In addition, it is noticed that $\Delta$\textsubscript{CT} is further reduced by 0.4 eV for $H$ = 6 T as compared to $H$ = 0, suggesting the increased contribution of hybridized \textit{d}\textsuperscript{n+1}$\underline{L}$ due to the enhanced covalency. The strengthening of Co 3$d$ - O 2$p$ hybridization is also manifested in the increase of $V(e_g)$ and $V(t_{2g})$. The changes observed in the energy difference $U_{dd}$-$U_{pd}$ (where $U_{pd}$ is the core-hole potential) are also expected due to the slight increase of electron occupation at Co for $H$ = 6 T. 

Contrasting with the results observed at the Co $L_{2,3}$-edge, the low temperature Mn $L_{2,3}$-edge XAS shows negligible variation with $H$ (see SM \cite{SM}). The O K-edge, on the other hand, exhibits an overall $H$-induced increase of its spectral weight on the pre-peak region \cite{SM}. These findings suggest that O-2$p$ electrons mainly contribute to the $H$-induced increase in the population of Co orbitals, since the depopulation of Mn orbitals is very subtle. 

The electronic parameters obtained in the multiplet calculations were used to estimate the ground state configurations of Co, using the software CTM4DOC \cite{CTM4DOC}. The resulting electronic configurations, depicted in Table \ref{tab:2}, evidence a subtle $H$-induced increase in the portion of the $t_{2g}$$^{5}e_g$$^2$ + $t_{2g}$$^{6}e_g$$^{2}\underline{L}$ states, naturally associated with the slight increase of 10Dq. At the same time, the increased orbital hybridization leads to significant enhancements in the amounts of ligand-field states $t_{2g}$$^{6}e_g$$^{2}\underline{L}$, $t_{2g}$$^{5}e_g$$^{3}\underline{L}$  and $t_{2g}$$^{4}e_g$$^{4}\underline{L}$.

\subsection{Density functional theory}
\label{DFT}

The above-described multiplet calculations yield valuable information regarding the temperature- and $H$-induced modulations of the 3$d$ electronic configurations in LCMO. However, it approximately assumes an $O_h$ crystal field on the O-octahedron, i.e. it neglects the lowered orthorhombic symmetry. Therefore, in order to get further insights from the real structure, we performed DFT calculations based on the XRD-delivered crystal structure.

\begin{figure*}
\begin{center}
\includegraphics[width= \textwidth]{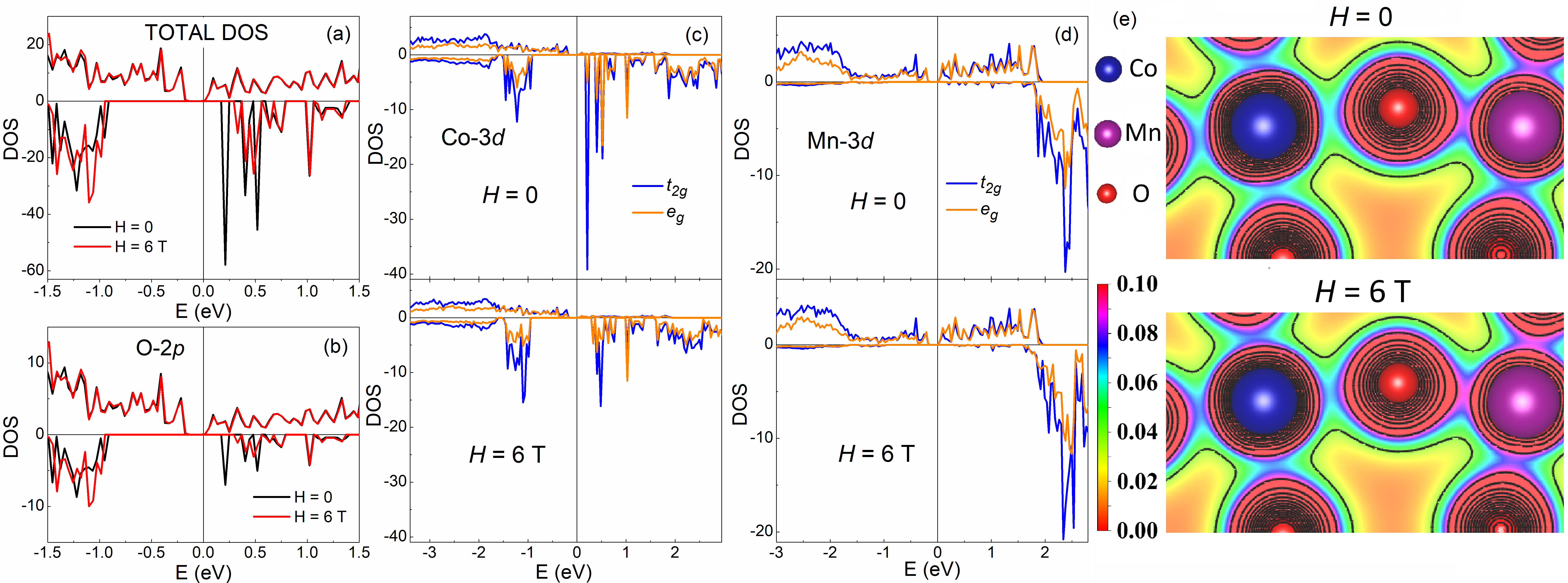}
\end{center}
\caption{(a) Total density of states for disordered LCMO with $H$ = 0 and $H$ = 6 T. (b) O-2$p$, (c) Co-3$d$ and (d) Mn-3$d$ orbital-resolved projected density of states for $H$ = 0 and 6 T. (e) Contour plots of the spatial charge density distribution around Co, Mn and O at $H$ = 0 and 6 T. The cutoff for the charge density contour was chosen to be 10\% of the maximum in each case.}
\label{Fig_DFT}
\end{figure*}

The DFT calculations on disordered LCMO was performed using an 8 f.u. SQS structure where four Co and four Mn atoms occupy their correct lattice sites whereas the remaining eight octahedral sites are occupied by an equal number of Co and Mn antisites, a protocol similar to that employed in other disordered DPs \cite{PRB2024}. The initial orthorhombic model was extracted from the 150 K XRD measurement carried out with $H$ = 0. To simulate the effect of $H$ = 6 T at low temperature, we have used the $\Delta L(H)$ measured at 15 K for $H$ = 6 T, and the $a$, $b$, $c$ lattice parameters used were calculated assuming that the proportional contribution of each lattice parameter to the $\Delta L(H)$ observed at 150 K holds at lower temperatures. Fig. \ref{Fig_DFT} shows the resulting density of states (DOS), where the Fermi energy ($E_F$) is set to zero, and Table \ref{T_DFT} displays the integrated DOS for each individual Co- and Mn-3$d$ orbital. 

Fig. \ref{Fig_DFT}(a) shows a magnified view of the total DOS calculated for $H$ = 0 and 6 T at the region around $E_F$, while \ref{Fig_DFT}(b) depicts the orbital-resolved projected DOS for the O 2$p$ orbitals (for a complete view of the DOS at a larger energy range, see SM \cite{SM}). The gap between the occupied states and the conduction band confirms the insulating character of LCMO \cite{Murthy3}. Both figures reveal a subtle decrease in the density of unoccupied states near $E_F$ when $H$ is applied, as well as the corresponding increase in the density of occupied levels. These changes support the scenario of $H$-induced increase of the TM-O hybridization. 

The orbital-resolved DOS for the $t_{2g}$ and $e_g$ levels of the Co-$3d$ orbitals are depicted in Fig. \ref{Fig_DFT}(c). The unoccupied states are mostly of spin-down character, as expected for a nearly Co$^{2+}$ (3$d^7$) state in HS configuration in which all the spin-up states are occupied. Comparing the DOS for $H$ = 0 to that of $H$ = 6 T, one can notice a clear decrease in the density of unoccupied states, while the density of occupied states increases. This is in agreement with a scenario of increased Co-O hybridization and O $\rightarrow$ Co charge-transfer, induced by the lattice contraction. Indeed, Table \ref{T_DFT} shows an overall increase in the occupation of both $e_g$ and $t_{2g}$ orbitals.

For the Mn-3$d$ orbitals, Fig. \ref{Fig_DFT}(d) shows that the DOS for $E$ $<$ $E_F$ almost entirely consists of spin-up states, while the unoccupied levels present both spin-up and spin-down characters. This is in agreement with a nearly Mn$^{4+}$ (3$d^3$) configuration. Comparing the results for $H$ = 0 and 6 T, the changes are much more subtle than those observed for Co. Again, this agrees with the experimental XAS data \cite{SM}. Overall, it is noticed a very subtle increase in the DOS at $H$ = 6 T compared to $H$ = 0, suggesting a slight enhancement of the Mn-O hybridization. 

Contour plots (isosurfaces) of the charge densities around Co, Mn and O for $H$ = 0 and 6 T are depicted in Fig. \ref{Fig_DFT}(e). The figure shows subtle changes in the spatial charge distribution, where increased charge densities can be seen in the regions at the immediacy of TM-O bonds, in special around Co. Such changes are expected to affect the polar regions responsible for the material's dielectric response, being thus in the core of the magnetodielectric effect observed on LCMO. 

\begin{table}
    \centering
    \begin{tabular}{c|c|c|c|c}
    \hline \hline
\multicolumn{5}{c}{Co-3$d$} \\ \hline
    
      &  \multicolumn{2}{c|}{$E$ $<$ $E_F$} & \multicolumn{2}{c}{$E$ $>$ $E_F$} \\ \hline  
    
    Orbital  & $H$ = 0  & $H$ = 6 T  & $H$ = 0  & $H$ = 6 T \\ \hline
      
   $d_{xz}$ & 11.59 & 12.49 & 3.18 & 1.81 \\
   
   $d_{yz}$ & 11.81 & 12.01 & 3.02 &  3.23 \\
   
   $d_{xy}$ & 10.30 & 10.73 & 4.61 & 4.61 \\
   
   $d_{z^2}$ & 11.16 & 13.07 & 3.43 & 2.96 \\
   
   $d_{x^2-y^2}$ & 10.62 & 11.32 & 4.13 & 4.32 \\

\hline

Total $t_{2g}$ & 33.70 & 35.23 & 10.81 & 9.65 \\

Total $e_g$ & 21.78 & 24.39 & 7.56 & 7.28 \\

\hline \hline
   
\multicolumn{5}{c}{Mn-3$d$} \\ \hline
    
       &  \multicolumn{2}{c|}{$E$ $<$ $E_F$} & \multicolumn{2}{c}{$E$ $>$ $E_F$} \\ \hline  
    
    Orbital  & $H$ = 0  & $H$ = 6 T  & $H$ = 0  & $H$ = 6 T \\ \hline
      
   $d_{xz}$ & 8.17 & 8.19 & 5.62 & 5.77 \\
   
   $d_{yz}$ & 8.01 & 8.06 & 5.66 & 6.03 \\
   
   $d_{xy}$ & 7.31 & 7.38 & 5.87 & 6.07 \\
   
   $d_{z^2}$ & 7.97 & 7.99 & 5.75 & 6.10 \\
   
   $d_{x^2-y^2}$ & 7.27 & 7.35 & 6.19 & 6.38 \\

\hline

Total $t_{2g}$ & 23.50 & 23.63 & 17.15 & 17.87 \\

Total $e_g$ & 15.24 & 15.34 & 11.94 & 12.47 \\
   
\hline \hline    
           
    \end{tabular}
    \caption{DFT-derived integrated DOS for Co- and Mn-3$d$ orbitals.}
    \label{T_DFT}
\end{table}

\subsection{Discussion}
\label{discussion}

LCMO is well known for its insulating character and FM nature, mostly ascribed to pseudo-linear superexchange interactions along the Co$^{2+}$-O-Mn$^{4+}$ path governed by the Goodenough-Kanamori-Anderson rules \cite{Blasse,Goodenough3}. Many physical properties of this system are particularly susceptible to the synthesis conditions \cite{Goodenough4}, since Co$^{2+}$/Co$^{3+}$ and Mn$^{4+}$/Mn$^{3+}$ mixed valence states are often present, giving rise to a second FM transition related to Co$^{3+}$--O--Mn$^{3+}$. This is the case of our LCMO sample, for which a previous study revealed a vast predominance of Co$^{2+}$/Mn$^{4+}$ ions, with the presence of a minor amount of Co$^{3+}$/Mn$^{3+}$ \cite{PRB2019}. The dielectric response of LCMO is usually ascribed to Co$^{2+}$/Mn$^{4+}$ charge ordering that leads to asymmetric polaronic hopping \cite{Kumar,Singh,Lin,Fournier,Murthy3}. In the context of the MD effect, our present study shows a giant and anisotropic $H$-induced lattice shrinkage below $T_C$, which in turn modulates the electronic structures of the TM ions and its hybridization with neighboring O ions, being this effect particularly relevant on Co.

\begin{figure}
 \centering
 \includegraphics[width=1\columnwidth]{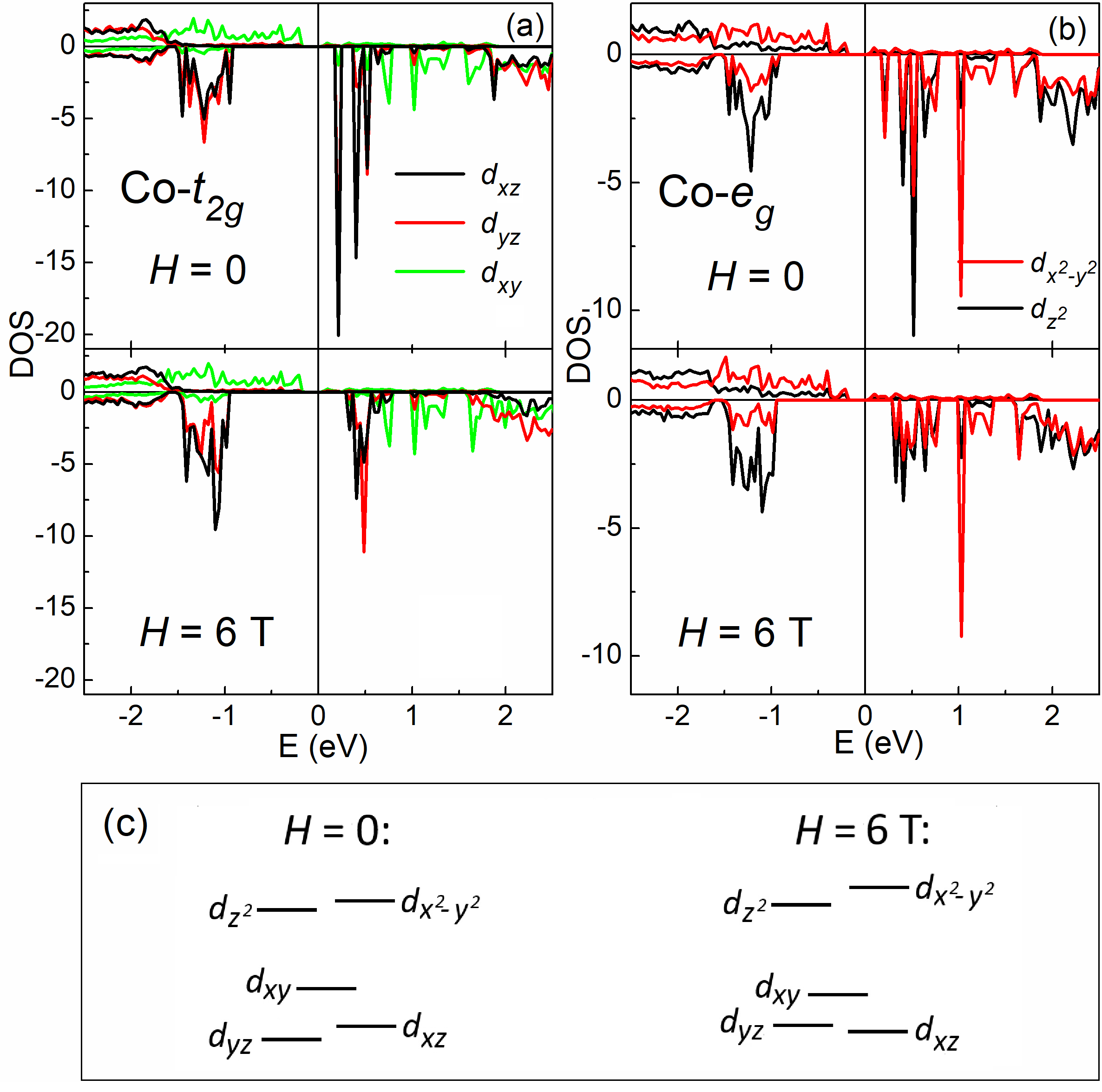}
\caption{DOS on Co-3$d$ (a) $t_{2g}$ and (b) $e_{g}$ orbitals, at $H$ = 0 and $H$ = 6 T. (b) Qualitative scheme of the energy levels for Co-3$d$ orbitals at $H$ = 0 and $H$ = 6 T.}
\label{Fig_draw}
\end{figure}

It is well known that the octahedral crystal field lifts the $d$ orbital degeneracy into the $t_{2g}$ and $e_g$ levels. In the case of LCMO, our XRD data evidences roughly similar TM-O bond lengths along the equatorial $ac$ plane, while a larger TM-O distance is observed along the apical $b$ parameter. From this, as a first step we can approximate the crystal field surrounding the Co ions to a tetragonally elongated octahedra that further lifts the $e_g$ and $t_{2g}$ degeneracies, with the $t_{2g}$ orbitals being separated into a $\langle L_z \rangle$ = 1 doublet lying in a lower energy level than a $\langle L_z \rangle$ = 0 singlet. This is in agreement with Table \ref{T_DFT}, as well as Fig. \ref{Fig_draw}(a) showing the DOS for each Co-$t_{2g}$ orbital calculated for $H$ = 0, at the region close to $E_F$. As the figure and the table show, the $d_{xy}$ orbital is less populated, suggesting that it corresponds to the $\langle L_z \rangle$ = 0 singlet at a higher energy level, while the more populated $d_{xz}$ and $d_{yz}$ orbitals form the lower energy doublet. However, our real system is not a purely tetragonally distorted octahedra. The XRD also revealed differences in the B--O lengths along $a$ and $c$, as well as tilts of the O-Co-O angles in the $ac$ and $ab$ planes, representing further distortions in the orthorhombic LCMO system that contribute to the redistribution of the energy levels \cite{Baidya}. Fig. \ref{Fig_draw}(a) and Table \ref{T_DFT} also show that $d_{yz}$ is slightly more populated than $d_{xz}$, which leaves the former in a lower energy level, a consequence of the fact that the $c$ lattice parameter is slightly larger than $a$. Fig. \ref{Fig_draw}(c) summarizes this scenario in a qualitative scheme for the energy levels of the Co-3$d$ orbitals.

It is interesting to note in Table \ref{T_DFT} that, albeit the application of $H$ leads to an overall decrease in the density of unoccupied states at Co-3$d$, some individual orbitals follow opposite trends. This suggests an internal redistribution of the electrons in the Co levels. The $d_{xz}$ orbital, initially slightly less populated than $d_{yz}$, undergoes a significant decrease in its density of unoccupied states with the application of $H$, whereas the opposite trend occurs for $d_{yz}$. This results in the inversion of the relative occupation of these states, likely because the $H$-induced lattice compression is more pronounced along $c$ ($y$) than along $a$ ($x$). Such changes in the relative occupancies of $d_{xz}$ and $d_{yz}$ may be the cause of the $H$-induced inversion in the spectral weights of the fairly close features B and C observed in the Co $L_3$-edge XAS (Fig. \ref{Fig_CoL_H}). 

Regarding the $e_g$ orbitals, the $d_{z^2}$ level is expected to be more affected by the anisotropic lattice compression along $b$ ($z$). Therefore, the density of its unoccupied states decreases, while that of $d_{x^2-y^2}$ increases. Accounting to the fact that the separation in energy between features D and B/C ($\sim$1 eV) is reasonably close to the 10Dq value obtained from multiplet calculations (which also agrees with resonant inelastic x-ray scattering on octahedraly coordinated Co \cite{Mattie,Groot1}), we conclude that the $H$-induced increase in the density of accessible states at the $d_{x^2-y^2}$ orbital may be responsible for the relative increase of feature D in the XAS data. 

Contrasting with the other orbitals, the occupancy of $d_{xy}$ remains nearly unaltered with the application of $H$. Here we recall that the magnetoelastic properties of Co-based materials were previously attributed to $H$-induced reorientation of spin and/or angular moments \cite{Manikandan,Slonczewski,Chai}. Within this scenario, given the unquenched orbital moment and the non-negligible SOC in octahedrally coordinated Co$^{2+}$, some rotation of the orbital moment is presumed to occur with the application of $H$, in addition to the tendency of the spin moment to align with the field. This is expected to be more pronounced on the $e_g$ levels as well as on the $t_{2g}$ orbitals presenting non-negligible angular moment, \textit{i.e.} the $\langle L_z \rangle$ = 1 $d_{xz}$+$d_{yz}$ orbitals, but not for the $\langle L_z \rangle$ = 0 $d_{xy}$ singlet. This explains the negligible modulation of the latter orbital with $H$. In this same context, Mn$^{4+}$ is also expected to be minimally affected by the field because, despite its significant spin moment, the similar occupation of its $t_{2g}$ orbitals results in a negligible angular momentum. Therefore, we conclude that the MD effect of LCMO is mostly originated from the $H$-induced redistribution of the electron occupation in Co-3$d$ energy levels, which in turn alters the spatial charge distribution around this polar region. 
 
\section{Summary}

In summary, we conducted a systematic investigation of the coupling between the structural, electronic and magnetic properties of a polycrystalline La$_2$CoMnO$_6$ perovskite. Several experimental techniques and ab initio calculations within the DFT framework were employed aiming to deeply comprehend the microscopic mechanism responsible for the material's magnetodielectric behavior. Our XRD data, combined with capacitive dilatometer measurements, reveal anisotropic $H$-induced lattice shrinkage responsible for a giant negative magnetostriction at low temperatures. In addition, the XAS measurement results, consistently supported by multiplet and DFT calculations, indicate that the $H$-induced increase in the crystal field strength modulates the Co-O orbital hybridization and charge transfer processes. This leads to the redistribution of the spatial electronic densities around this ion, therefore altering the dielectric response of the material. Our findings provide valuable insights into the behavior and physical properties of a magnetodielectric double perovskite oxide and, furthermore, contribute to a better understanding of its potential for technological applications.

\begin{acknowledgements}

This work was supported by the Brazilian funding agencies: Funda\c{c}\~{a}o Carlos Chagas Filho de Amparo \`{a} Pesquisa do Estado do Rio de Janeiro (FAPERJ) [Nos. E-26/204.206/2024 and E-26/211.291/2021], Funda\c{c}\~{a}o de Amparo \`{a}  Pesquisa do Estado de Goi\'{a}s (FAPEG) and Conselho Nacional de Desenvlovimento Cient\'{\i}fico e Tecnol\'{o}gico (CNPq) [Grants 400633/2016-7, 309599/2021-0, 305394/2023-1, 303772/2023-9, 405408/2023-4]. R.B.P. acknowledges LaMCAD-UFG, supercomputer SDumont/LNCC-MCTI and Cluster Euler/CeMEAI for providing the computational resources. M. de S. acknowledges partial financial support from the S\~ao Paulo Research Foundation - Fapesp (Grants 2011/22050-4, 2017/07845-7, and 2019/24696-0). R.L.S. acknowledges FAPEMIG-MG (grant APQ-02378-17L). L.S. acknowledges IGCE for the post-doc fellowship. We thank the XDS staff of LNLS for technical support and LNLS for the concession of beam time (proposals Nos. 20170612 and 20180745). We also thank Diamond Light Source for time on beamline I06 under proposal MM35100-1. 

\end{acknowledgements}


\begin{thebibliography}{99}

\bibitem{Anderson} P. W. Anderson, New Approach to the Theory of Superexchange Interactions, Phys. Rev. \textbf{115}, 2 (1959).

\bibitem{Goodenough1} J. B. Goodenough, Phys. Rev. \textbf{100},Theory of the Role of Covalence in the Perovskite-Type Manganites [La,M(II)]O$_3$, Phys. Rev. \textbf{100}, 564 (1955).

\bibitem{Goodenough2} J. B. Goodenough, An interpretation of the magnetic properties of the perovskite-type mixed crystals La$_{1-x}$Sr$_x$Co0$_{3-\lambda}$, J. Phys. Chem. Solids \textbf{6}, 287 (1958).

\bibitem{Kanamori} J. Kanamori, Superexchange interaction and symmetry properties of electron orbitals, J. Phys. Chem. Solids \textbf{10}, 87 (1959).

\bibitem{Blasse} G. Blasse, Ferromagnetic interactions in non-metallic perovskites, J. Phys. Chem. Solids \textbf{26}, 1969 (1965).

\bibitem{Goodenough3} J. B. Goodenough, Relationship Between Crystal Symmetry and Magnetic Properties of Ionic Compounds Containing Mn$^{3+}$, Phys. Rev. \textbf{124}, 373 (1961).

\bibitem{Goodenough4} R. I. Dass and J. B. Goodenough, Multiple magnetic phases of La$_2$CoMnO$_{6-\delta}$ (0 $\leq\delta\leq$ 0.05),  Phys. Rev. B.  \textbf{67}, 014401 (2003).

\bibitem{Goodenough5} R. I. Dass, J.-Q. Yan, and J. B. Goodenough, Oxygen stoichiometry, ferromagnetism, and transport properties of La$_{2-x}$NiMnO$_{6+\delta}$, Phys. Rev. B \textbf{68}, 064415 (2003).

\bibitem{PRB2019} L. T. Coutrim, D. Rigitano, C. Macchiutti, T. J. A. Mori, R. Lora-Serrano, E. Granado, E. Sadrollahi, F. J. Litterest, M. B. Fontes, E. Baggio-Saitovitch, E. M. Bittar, and L. Bufai\c{c}al.   Zero-field-cooled exchange bias effect in phase-segregated  La\textsubscript{2-x}A\textsubscript{x}CoMnO\textsubscript{6-$\delta$}  (A = Ba, Ca, Sr; x = 0,5).  Phys. Rev. B.  \textbf{100}, 054428 (2019).

\bibitem{Singh} M. P. Singh, K. D. Truong, and P. Fournier, Magnetodielectric effect in La$_2$CoMnO$_6$ double perovskite thin films, Appl. Phys. Lett. \textbf{91}, 042504 (2007).

\bibitem{Sarma} D. Choudhury, P. Mandal, R. Mathieu, A. Hazarika, S. Rajan, A. Sundaresan, U. V. Waghmare, R. Knut, O. Karis, P. Nordblad, and D. D. Sarma, Near-Room-Temperature Colossal Magnetodielectricity and Multiglass Properties in Partially Disordered La$_2$NiMnO$_6$, Phys. Rev. Lett. \textbf{108}, 127201 (2012).

\bibitem{Hill} N. A. Hill, Why Are There so Few Magnetic Ferroelectrics?, J. Phys. Chem. B \textbf{104}, 29 (2000).

\bibitem{Fiebig} M. Fiebig, T. Lottermoser, D. Meier, and M. Trassin, The evolution of multiferroics, Nat. Rev. Mater. \textbf{1}, 16046 (2016).

\bibitem{Liu} H. Liu and X. Yang, A brief review on perovskite multiferroics, Ferroelectrics \textbf{507}, 69 (2017). 

\bibitem{Wang} J. Wang, J. B. Neaton, H. Zheng, V. Nagarajan, S. B. Ogale, B. Liu, D. Viehland, V. Vaithyanathan, D. G. Schlom, U. V. Waghmare, N. A. Spaldin, K. M. Rabe, M. Wuttig,  and R. Ramesh, Epitaxial BiFeO$_3$ Multiferroic Thin Film Heterostructures, Science \textbf{299}, 5613 (2003).

\bibitem{Zhu} Ch. Jooss, L. Wu, T. Beetz, R. F. Klie, M. Beleggia, M. A. Schofield, S. Schramm, J. Hoffmann, and Y. Zhu, Polaron melting and ordering as key mechanisms for colossal resistance effects in manganites, Proc. Natl Acad. Sci. USA \textbf{104}, 34 (2007).

\bibitem{Nagaosa} Y. Tokura, S. Seki, and N. Nagaosa, Multiferroics of spin origin, Rep. Prog. Phys. \textbf{77}, 076501 (2014).

\bibitem{Kimura} T. Kimura, T. Goto, H. Shintani, K. Ishizaka, T. Arima, and Y. Tokura, Magnetic control of ferroelectric
polarization, Nature \textbf{426}, 55-58 (2003).
 
\bibitem{Lin} Yi Qi Lin and Xiang Ming Chen, Dielectric, Ferromagnetic Characteristics, and Room Temperature Magnetodielectric Effects in Double Perovskite La$_2$CoMnO$_6$ Ceramics, J. Am. Ceram. Soc., \textbf{94} [3] 782 (2011).

\bibitem{Murthy3} J. Krishna Murthy, K. D. Chandrasekhar, S. Murugavel, and A. Venimadhav, Investigation of the intrinsic magnetodielectric effect in La$_2$CoMnO$_6$: role of magnetic disorder, J. Mater. Chem. C \textbf{3}, 836 (2015).

\bibitem{Blasco} A. J. Bar\'{o}n-Gonz\'{a}lez, C. Frontera, J. L. Garc\'{i}a-Mu\~{n}oz, B. Rivas-Murias and J. Blasco, Effect of cation disorder on structural, magnetic and dielectric properties of La$_2$MnCoO$_6$ double perovskite, J. Phys.: Condens. Matter \textbf{23}, 496003 (2011).

\bibitem{Kumar} K. Manna, R. S. Joshi, S. Elizabeth, and P. S. Anil Kumar, Evaluation of the intrinsic magneto-dielectric coupling in LaMn$_{0.5}$Co$_{0.5}$O$_3$ single crystals, Appl. Phys. Lett. \textbf{104}, 202905 (2014).

\bibitem{SM} See Supplemental Material at [URL] for details of the crystal growth and additional results regarding the physical properties of LCMO.

\bibitem{XDS} F. A. Lima, M. E. Saleta, R. J. S. Pagliuca, M. A. Eleot\'{e}rio, R. D. Reis, J. Fonseca J\'{u}nior, B. Meyer, E. M. Bittar, N. M. Souza-Neto, and E. Granado, XDS: a flexible beamline for X-ray diffraction and spectroscopy at the Brazilian synchrotron, J. Synchrotron Radiat. \textbf{23}, 1538 (2016).

\bibitem{GSAS} A. C. Larson and R. B. Von Dreele, Los Alamos National Laboratory Report No. LAUR 86-748, 2000; B. H. Toby, J. Appl. Crystallogr. \textbf{34}, 210 (2001).

\bibitem{Crispy} Marius Retegan, Stephan Kuschel. Crispy: v0.7.4 (2019), DOI 10.5281/zenodo.1008184.

\bibitem{CTM4DOC} Mario Ulises Delgado-Jaime, Kaili Zhang, Josh Vura-Weis and Frank M. F. de Groot. CTM4DOC: electronic structure analysis from X-ray spectroscopy, J. Synchrotron Rad. \textbf{23}, 1264-1271 (2016).

\bibitem{Kresse} G. Kresse and J. Furthm\"{u}ller. Efficient iterative schemes for ab initio total-energy calculations using a plane-wave basis set, Phys. Rev. B, \textbf{54}, 11169-11186 (1996). 

\bibitem{Kresse2} G. Kresse and J. Furthm\"{u}ller. Efficiency of ab-initio total energy calculations for metals and semiconductors using a plane-wave basis set, Comput. Mater. Sci.  \textbf{6}, 15-50 (1996).

\bibitem{Hohenberg} P. Hohenberg and W. Kohn. Inhomogeneous Electron Gas, Phys. Rev. B \textbf{136}, 864, (1964). 

\bibitem{Kohn} W. Kohn and L. Sham. Self-Consistent Equations Including Exchange and Correlation Effects, Phys. Rev.  \textbf{140}, A1133 (1965). 

\bibitem{Perdew} J. P. Perdew, K. Burke, and  M. Ernzerhof. Generalized Gradient Approximation Made Simple, Phys. Rev. Lett. \textbf{77}, 3865-3868 (1996).

\bibitem{Kresse3} G. Kresse and D. Joubert. From ultrasoft pseudopotentials to the projector augmented-wave method, Phys. Rev. B \textbf{59}, 1758-1775 (1999).
 
\bibitem{Monkhorst} H. J. Monkhorst and J. D. Pack. Special points for brillouin-zone integrations, Phys. Rev. B \textbf{13}, 51885192 (1976). 

\bibitem{Wei} S.-H. Wei, L. G. Ferreira, J. E. Bernard, and Alex Zunger. Electronic properties of random alloys: Special quasirandom structures, Phys. Rev. B \textbf{42}, 9622 (1990).

\bibitem{Walle}	A. van de Walle, M. Asta, G. Ceder. The Alloy Theoretic Automated Toolkit: A user guide, Calphad \textbf{26}, 539 (2002).

\bibitem{Padilha} J. E. Padilha, L. Seixas, R. B. Pontes, A. J. R. da Silva, and A. Fazzio. Quantum spin Hall effect in a disordered hexagonal Si$_x$Ge$_{1-x}$ alloy, Phys. Rev. B \textbf{88}, 201106(R) (2013).

\bibitem{Rashid} M. Rashid, N. A. Noor, B. Sabir, S. Ali, M. Sajjad, F. Hussain, N. U. Khan, B. Amin, R. Khenata. Ab-initio study of fundamental properties of ternary ZnO$_{1-x}$S$_x$ alloys by using special quasi-random structures, Comput. Mater. Sci. \textbf{91}, 285-291 (2014). 

\bibitem{Fournier} K. D. Truong, J. Laverdi\`{e}re, M. P. Singh, S. Jandl, and P. Fournier, Impact of Co/Mn cation ordering on phonon anomalies in La2CoMnO6 double perovskites: Raman spectroscopy, Phys. Rev. B \textbf{76}, 132413 (2007).

\bibitem{Joy} P. A. Joy, Y. B. Khollam, and S. K. Date, Spin states of Mn and Co in LaMn$_{0.5}$Co$_{0.5}$O$_3$, Phys. Rev. B \textbf{62}, 8608 (2000).

\bibitem{Joly} V. L. Joseph Joly, P. A. Joy, S. K. Date and C. S. Gopinath, The origin of ferromagnetism in the two different phases of LaMn$_{0.5}$Co$_{0.5}$O$_3$: evidence from x-ray photoelectron spectroscopic studies, J. Phys.: Condens. Matter \textbf{13} 649 (2001).

\bibitem{Bull} C. L. Bull, H. Y. Playford, K. S. Knight, G. B. G. Stenning, and M. G. Tucker. Magnetic and structural phase diagram of the solid solution LaCo$_x$Mn$_{1-x}$O$_3$, Phys. Rev. B \textbf{94}, 014102 (2016). 

\bibitem{Manikandan} M. Manikandan, A. Ghosh, R. Mahendiran. Giant magnetostriction in La$_2$CoMnO$_6$ synthesized by microwave irradiation, Appl. Phys. Lett. \textbf{123}, 022403 (2023).

\bibitem{mir} L. Lopez-Mir, R. Galceran, J. H. Martin, B. Bozzo, J. C. Fernanadez, E. V. P. Miner, A. Pomar, L. Balcells, B. Martinez, and C. Frontera. Magnetic anisotropy and valence states in  La\textsubscript{2}Co\textsubscript{1-x}Mn\textsubscript{1+x}O\textsubscript{6} (x $\approx$ 0.23) thin-films studied by x-ray absorption spectroscopy techniques. Phys. Rev. B. \textbf{95}, 224434 (2017).

\bibitem{Asai} Kichizo Asai, Atsuro Yoned1, Osamu Yokokura, J. M. Tranquada, G. Shirane, and Key Kohn. Two Spin-State Transitions in LaCoO$_3$, J. Phys. Soc. Jpn. \textbf{67}, 290-296 (1998).

\bibitem{Raveau} B. Raveau and Md. Motin Seikh. Cobalt Oxides: From Crystal Chemistry to Physics (Wiley-VCH, Weinheim, 2012).

\bibitem{zaanen}  J. Zaanen, G. A. Sawatzky, and J. W. Allen, Band gaps and electronic structure of transition-metal compounds. Phys. Rev. Lett.  \textbf{55}, 419-421 (1985).

\bibitem{Harrison} W. A. Harrison, Electronic Structure and Physical Properties of Solids (Freeman, San Francisco, 1980). 


\bibitem{Jana} A. Jana, R. J. Choudhary, D. M. Phase . Mott-Hubbard type insulating nature of epitaxial LaVO\textsubscript{3} thin films. Phys. Rev. B.  \textbf{98}, 075124 (2018).

\bibitem{Bagri} A. Bagri, S. Sahoo, R. J. Choudhary, D. M. Phase, Electronic manifestation of a disorder mediated metal-insulator tranistion in epitaxial SrRuO\textsubscript{3} thin film. J. Alloys. Compd.  \textbf{902}, 163644 (2022).

\bibitem{PRB2024} A. G. Silva, R. B. Pontes, M. Boldrin, H. V. S. Pessoni, L. S. I. Veiga, J. R. Jesus, H. Fabrelli, A. R. C. Gonzaga, E. M. Bittar, and L. Bufai\c{c}al, Near-room-temperature ferrimagnetism and half-metallicity in disordered Ca$_{1.5}$La$_{0.5}$MnRuO$_6$. Phys. Rev. B.  \textbf{110}, 144415 (2024).

\bibitem{Baidya} S.Baidya, T. Saha-dasgupta. Electronic structure and phonons in La\textsubscript{2}CoMnO\textsubscript{6}: A ferromagnetic insulator driven by Coulomb-assisted spin-orbit coupling, Phys. Rev. B \textbf{84}, 035131 (2011).

\bibitem{Mattie} M. M. an Schooneveld, R. Kurian, A. Juhin, K. Zhou, J. Schlappa, V. N. Strocov, T. Schmitt, and F. M. F. de Groot. Electronic structure of CoO nanocrystals and a single crystal probed by resonant x-ray emission spectroscopy. J. Phys. Chem. C.  \textbf{116}, 15218-15230 (2012).

\bibitem{Groot1} Ru-Pan Wang, Boyang Liu, Robert J. Green, Mario Ulises Delgado-Jaime, Mahnaz Ghiasi, Thorsten Schmitt, Matti M. van Schooneveld, and Frank M. F. de Groot, Charge-Transfer Analysis of 2p3d Resonant Inelastic X-ray Scattering of Cobalt Sulfide and Halides. J. Phys. Chem. C \textbf{121}, 24919-24928 (2017).

\bibitem{Slonczewski} J. C. Slonczewski. Origin of Magnetic Anisotropy in Cobalt-Substituted Magnetite, Phys. Rev. \textbf{110}, 1341 (1958).

\bibitem{Chai} Yi-Sheng Chai, Jun-Zhuang Cong, Jin-Cheng He, Dan Su, Xia-Xin Ding, John Singleton, Vivien Zapf, and Young Sun. Giant magnetostriction and nonsaturating electric polarization up to 60 T in the polar magnet CaBaCo$_4$O$_7$, Phys. Rev. B \textbf{10}, 174433 (2021).

\bibitem{Groot2} Mario Ulises Delgado-Jaime, Kaili Zhang, Josh Vura-Weis and Frank M. F. de Groot, CTM4DOC: electronic structure analysis from X-ray spectroscopy. J. Synchrotron Rad. \textbf{23}, 1264-1271 (2016).

\end{thebibliography}
\end{document}